\documentclass[aps, prl,twocolumn, showpacs, superscriptaddress]{revtex4-1}
\usepackage{graphicx}
\usepackage{amsmath}
\usepackage{amsfonts}
\usepackage{amssymb}
\usepackage{hyperref}

\newcommand{\beq}{\begin{equation}}
\newcommand{\eeq}{\end{equation}}

\begin{document}

\title{Prethermalization of atoms due to photon-mediated long-range interactions }

\author{Stefan Sch\"utz} 
\affiliation{Theoretische Physik, Universit\"at des Saarlandes, D-66123 Saarbr\"ucken, Germany} 

\author{Giovanna Morigi} 
\affiliation{Theoretische Physik, Universit\"at des Saarlandes, D-66123 Saarbr\"ucken, Germany} 

\date{\today}

\begin{abstract}
Atoms can spontaneously form spatially-ordered structures  in optical resonators when they are transversally driven by lasers. This occurs when the laser intensity exceeds a threshold value and results from the mechanical forces on the atoms associated with superradiant scattering into the cavity mode. We treat the atomic motion semiclassically and show that, while the onset of spatial ordering depends on the intracavity-photon number, the stationary momentum distribution is a Maxwell-Boltzmann whose width is determined by the rate of photon losses. Above threshold, the dynamics is characterized by two time scales: after a violent relaxation, the system slowly reaches the stationary state over time scales exceeding the cavity lifetime by several orders of magnitude. In this transient regime the atomic momenta form non-Gaussian metastable distributions, which emerge from the interplay between the long-range dispersive and dissipative mechanical forces of light. We argue that the dynamics of selforganization of atoms in cavities offers a testbed for studying the statistical mechanics of long-range interacting systems.
\end{abstract}

\pacs{37.30.+i, 42.65.Sf, 05.65.+b, 05.70.Ln}

\maketitle

Long-range interactions characterize the dynamics of systems from microscopic to macroscopic scales, ranging from nuclear to astrophysical distances  \cite{Campa:2009}. In these systems the individual components can interact with a long-range potential that decays with the interparticle distance $r$ slower than $r^{-d}$  in $d$ dimensions. This property leads, to mention some, to ensemble inequivalence and to the existence of quasi-stationary states, i.e., metastable states with non-thermal distributions \cite{Campa:2009}. 

\begin{figure}[hbt]
\begin{center}
\includegraphics[width=0.35\textwidth]{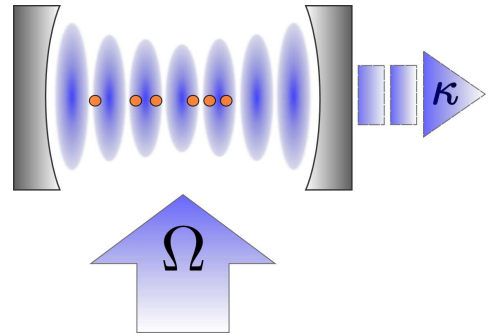}
\caption{\label{fig:1} (color online) Atoms in a standing-wave cavity and driven by a transverse laser can spontaneously form ordered patterns when the laser intensity $\Omega$ exceeds the rate of photon losses, here due to cavity decay at rate $\kappa$. In this regime the atoms experience a long-range interaction mediated by the cavity photons and their motion becomes strongly correlated. }
\end{center}
\end{figure}

Cold atoms driven by laser light constitute a promising laboratory realization of long-range interacting systems \cite{Cataliotti,RMP:2013,ODell,Labeyrie}. Here, multiple scattering of photons by atoms gives rise to mechanical forces which are infinitely long ranged when the atoms couple to a single-mode high-finesse cavity  \cite{Rempe}. In the overdamped regime this long-ranged potential lies at the origin of synchronization \cite{JILA} and collective atomic recoil lasing \cite{CARL_Zimmerman}.  When the cavity mode is a standing wave and the atoms are transversally pumped, as in the setup sketched in Fig. \ref{fig:1}, spontaneous ordering in spatially-periodic structures occurs \cite{Domokos:2002,Black:2003,Baumann:2010,RMP:2013}.  The phenomenon can be described in terms of formation of atomic gratings which maximize coherent scattering of laser photons into the cavity mode. These "Bragg gratings" are stably trapped by the  mechanical effects of the light they scatter, provided that the laser compensates the cavity losses so that the number of intracavity photons is sufficiently large. This takes place when the  strength of the laser coupling exceeds a threshold value $\Omega_c$ depending, amongst others, on the rate of photon losses and the number of atoms $N$ that couple with the cavity mode \cite{Asboth:2005,Niedenzu:2012}. This spatial selforganization was first predicted in Refs. \cite{Domokos:2002,ODell} and then reported in a series of experiments at laser-cooling temperatures \cite{Black:2003,Arnold:2012} and in the ultracold regime \cite{Baumann:2010,Mottl:2012}.

In this Letter we theoretically analyse the dynamics leading to the formation of spatial structures and their stationary properties in one dimension. For this purpose we resort to a Fokker-Planck equation (FPE) derived when the atoms are classically polarizable particles, their center-of-mass motion is treated semiclassically, while the cavity field is a full quantum variable \cite{Schuetz:2013}. This semiclassical limit can be applied when the cavity linewidth $\kappa$ (which determines the scattering cross section) exceeds the recoil energy $\omega_r=\hbar k^2/(2m)$, scaling the exchange of mechanical energy between an atom of mass $m$ and a photon of wave number $k$. Our approach complements the one applied in  Refs. \cite{Horak:2001,Domokos:2002,Asboth:2005,Niedenzu:2012}, based on the assumption that the cavity field is a semiclassical variable. By treating the cavity field quantum mechanically, we determine its state for any value of the laser amplitude and in particular at threshold, where quantum fluctuations are important. This information is extracted provided that retardation effects in the scattering processes are perturbations, such that at leading order the field is determined by the instantaneous atomic distribution \cite{Dalibard:1989}. Thus, for $N$ identical atoms confined in one dimension along the cavity axis, the total scattering amplitude depends on their positions $x_1,\ldots,x_N$ within the cavity standing wave $\cos(kx)$ and the cavity electric field at time $t$ is $E_c(t)\propto \sqrt{N\bar n}\langle\Theta\rangle_t$. Here, $\bar n$ is the maximum intracavity-photon number per atom, and is thus controlled by the strength of the external laser pump \cite{footnote:1}, while the order parameter
$$\Theta=\sum_{j=1}^N\cos(kx_j)/N$$
characterizes spatial ordering in the cavity \cite{Asboth:2005}. The field  reaches its maximum when $ |\Theta| =1$, namely, when the atoms form a Bragg grating. The corrections to $E_c$ due to the atomic motion are systematically included in the following as perturbation, assuming  that the atoms Doppler shifts are smaller than the cavity linewidth $\kappa$ \cite{Schuetz:2013}. 

The averages $\langle \cdot\rangle_t$ are taken over the normalized distribution  $f(x_1,p_1;\ldots;x_N,p_N;t)$ at time $t$, where $p_1,\ldots,p_N$ are the atomic momenta and $f$ obeys the FPE \cite{Schuetz:2013} 
\begin{align}
\label{FPE}
&\partial_t f+\{f,H\}\simeq\\
&- \bar n\Gamma\sum_i \sin(kx_i) \partial_{p_i}\frac{1}{N}\sum_{j}\sin(kx_j) \left( p_j + \frac{m}{ \beta} \partial_{p_j}\right)f\,.\nonumber
\end{align}
Here, the left-hand side (LHS) contains the Poisson brackets with the Hamiltonian $H$ governing the coherent dynamics, that originate from the conservative mechanical forces of light. The right-hand side (RHS) contains the friction coefficient due to retardation and the diffusion, due to fluctuations of the cavity field  because of photon losses \cite{Murr:2006}: These terms are scaled by $\bar n$  and by the rate $\Gamma = 8\omega_r\kappa\Delta_c/(\Delta_c^2 + \kappa^2)$, with $\Delta_c=\omega_L-\omega_c$ the detuning between laser and cavity-mode frequencies, such that $\bar n\Gamma$ is the maximum damping rate of a single atom ($N=1$). In addition, $\hbar \beta=-4\Delta_c/(\Delta_c^2+\kappa^2)$. The Hamiltonian 
\begin{eqnarray}
\label{Hmf}
H=\sum_j \frac{p_j^2}{2m}+  \hbar\Delta_c\bar n N\Theta^2+{\rm O}(U)
\end{eqnarray}
contains the cavity-mediated potential, which scales with $\bar n$ and is attractive when $\Delta_c$ is negative. Hence, this detuning determines whether the formation of Bragg gratings is energetically favoured. Equation \eqref{Hmf} summarizes in a compact way a property which was observed in several previous works  \cite{Domokos:2002,Black:2003,Asboth:2005}. It is reported at leading order in $|NU/\Delta_c|$, where $U$ is the dynamical Stark shift due to the coupling with the cavity field \cite{footnote:1}, and whose effect is systematically included in the numerical simulations. 

Remarkably, at leading order in $|NU/\Delta_c|$ Eq. \eqref{Hmf} allows one to draw a direct connection with the Hamiltonian Mean Field (HMF) model, the workhorse of the statistical mechanics of systems with long-range interaction, which in a canonical ensemble exhibits a second-order phase transition from a paramagnetic to a ferromagnetic phase controlled by the temperature \cite{Campa:2009}. This analogy becomes explicit writing $\Theta^2=\sum_{i,j}(\cos(k(x_i + x_j))+\cos(k(x_i - x_j)))/(2N^2)$, which shows that $H$ is extensive as it satisfies Kac prescription  \cite{Campa:2009}, and suggests to identify $\Theta$ with the $x$-component of a two-dimensional magnetization. 

Differing from the HMF model, the term $\cos(k(x_i + x_j))$ originates from the underlying cavity standing-wave potential that breaks continuous translational invariance. Moreover,  the cavity coupling at higher order in $|NU/\Delta_c|$ gives rise to deviations from the Hamiltonian dynamics due to further terms in the LHS of Eq. \eqref{FPE} (see, e.g., \cite{Fernandez-Vidal:2010}) which are responsible for bistable behaviour \cite{Stamper-Kurn}. Retardation effects and cavity losses, in addition, can establish long-range correlations between the atoms, as visible by inspecting the RHS. In fact, diffusion is here due to global quenches of the cavity potential. Similarly, retardation effects modify the cavity potential \cite{Domokos:2003}.  When the density is uniform, the terms in the RHS reduce to the Langevin terms of a FPE which fulfills detailed balance and the model is analogous to the Brownian Mean Field model \cite{Chavanis}.  However, this is valid at all times only well below the selforganization threshold. Indeed, the stationary density is here controlled by $\bar n$, and thus by the laser intensity, which scales both the strength of the long-range coherent and incoherent forces. This becomes evident when studying the dynamics at the asymptotics: A solution of  $\partial_tf_{\infty}=0$ is the thermal distribution
$f_{\infty}=f_0\exp(-\beta H)$ for $\Delta_c<0$, with  $f_0$ normalizing factor. The temperature is  {\it independent} of the laser intensity and its minimum $k_B T_{\rm min}=\hbar\kappa/2$ is achieved for $\Delta_c=-\kappa$, as also found in Ref. \cite{Asboth:2005,Niedenzu:2012,Piazza:2014}  using different approaches. In \cite{Niedenzu:2012} the selforganization threshold $\bar n_c=(1+\kappa^2/\Delta_c^2)/4$ was estimated by means of a kinetic theory based on treating the cavity field semiclassically. This value is consistent with our results. 

\begin{widetext}
\begin{figure*}[hbt]
\begin{center}
\includegraphics[width=0.99\textwidth]{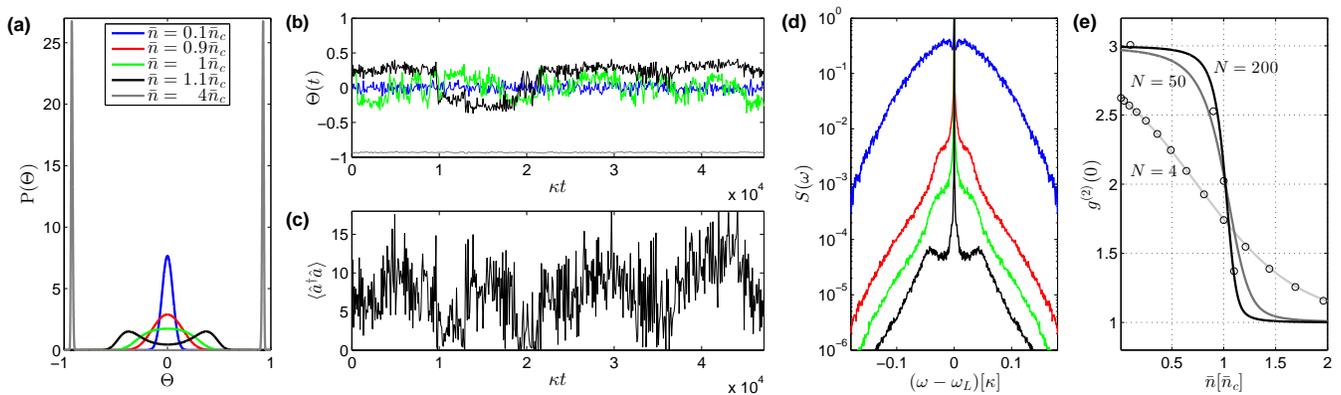}
\caption{\label{fig:2} (color online) (a) Distribution $P(\Theta)$ of the magnetization $\Theta$ at steady state for $\bar n/\bar n_c=0.1,\,0.9,\,1\,,1.1\,,4$ (see box for color code).  (b) Typical trajectories at the asymptotics for $N=200$ atoms are shown in (b) as a function of time (in units of $1/\kappa$) and for $\bar n/\bar n_c=0.1,\,1\,,1.1\,,4$. Note that the cavity field amplitude is proportional to $\Theta$. (c) Mean intracavity-photon number as a function of time for the trajectory at $\bar n=1.1\bar n_c$. (d) Spectrum $S(\omega)$ of the intensity of the emitted light (in arbitrary units) as a function of $\omega$ (in units of $\kappa$) for $\bar n/\bar n_c=0.1,\,0.9,\,1\,,1.1$ (from top to bottom).  (e) $g^{(2)}(0)$ as a function of $\bar n$ for different atoms numbers. The dots correspond to numerical results obtained by integrating the SDE. The cavity parameters are rescaled with $N$ so that $\bar n_c$ is independent on $N$ and finite (see \cite{Fernandez-Vidal:2010}). The atomic transition is the  $\text{D}_2$-line of $^{85}\text{Rb}$ at half linewidth $\gamma=2\pi\times 3$ MHz. The laser detuning from the atomic frequency is $\Delta_a = - 500\gamma$. Here, $\Delta_c=-\kappa$ with $\kappa=0.5\gamma$. 
}
\end{center}
\end{figure*}
\end{widetext}

We first discuss the predictions of Eq. \eqref{FPE} at the asymptotics. Figure \ref{fig:2}(a) displays the stationary distribution of the magnetization, $P(\Theta_0)=\langle\delta(\Theta_0-\Theta)\rangle_{\infty}$, for different values of $\bar n$. For $\bar n<\bar n_c$, $P(\Theta_0)$ is approximately a Gaussian centered at zero. At threshold it broadens and becomes increasingly localized at the values $\pm 1$ as $\bar n$ grows. The width of this distribution is determined by the fluctuations of the trajectories $\Theta(t)$: the larger $\bar n$ is, the more localized are the atoms at a Bragg grating, while the probability of a jump between gratings vanishes accordingly. Typical trajectories $\Theta(t)$ at the asymptotics of the dynamics are shown in Fig. \ref{fig:2}(b): They are obtained by integrating the stochastic differential equations (SDE) derived from Eq. \eqref{FPE} \cite{Schuetz:2013}. While below threshold $\Theta(t)$ fluctuates about zero (corresponding to a uniform spatial distribution), as $\bar n$ is increased above threshold it takes either positive or negative values, in which it remains trapped for time intervals which grow with $\bar n$. Jumps between the two values correspond to quenches of the intracavity-photon number following losses, as shown in (c) for $\bar n=1.1\bar n_c$, and take place over time intervals approximately scaling with the recoil frequency. Note that these jumps correspond to a simultaneous jump of all atomic trajectories out of the Bragg gratings \cite{Domokos:2002,Asboth:2005}. For $\bar n=4\bar n_c$ the residence time is infinite: photon losses give rise to small fluctuations of the potential depth and the atoms remain locked in a Bragg grating. These features determine the light amplitude at the cavity output, the jumps correspond to jumps of the field phase and can be measured by heterodyne detection \cite{Black:2003,Baumann:2011}. Additional information is contained in the power spectrum of the light intensity, which is  the Fourier transform $S(\omega)$ of the correlation function $g^{(1)}(\tau)=\lim_{t\to\infty}\langle \Theta(\tau+t)\Theta(t)\rangle/\langle |\Theta(t)|\rangle^2$ and is displayed in Fig. \ref{fig:2}(d) for different values of $\bar n$. $S(\omega)$ exhibits a narrow peak at the laser frequency  as the threshold is approached, and is associated with the creation of Bragg gratings coherently scattering light into the resonator.  The broad background spectrum is progressively suppressed, corresponding to a suppression of fluctuations of the order parameter as the atoms become localized in Bragg gratings. Moreover, at threshold two broad sidebands appear whose maximum moves away from $\omega=\omega_L$ as $\bar n$ increases from $\bar n_c$. A qualitative analysis shows that the sidebands width decreases as $\bar n$ is increased from $\bar n_c$. Similar features have been observed in the ultracold \cite{Mottl:2012,Baumann:2011,PNAS:2013} and have been interpreted in terms of density waves which drive the instability.  Figure \ref{fig:2}(e) displays the second-order correlation function of the emitted light at zero-time delay $g^{(2)}(0)$ as a function of $\bar n$, where $g^{(2)}(\tau)=\lim_{t\to\infty}\langle \Theta(\tau+t)^2\Theta(t)^2\rangle/\langle \Theta(t)^2\rangle^2$. Below threshold $g^{(2)}(0)\to 3$. This value is also found analytically after discarding correlations between the atoms. It monotonously decreases with $\bar n$ and reaches unity above threshold, $g^{(2)}(0)\to 1$, corresponding to a coherent state inside the resonator \cite{Habibian:2011}. The crossover between these two regimes narrows as the number of atoms is increased, suggesting a jump at $\bar n_c$ in the thermodynamic limit (here consisting in keeping $\bar n_c$ constant as $N\to\infty$ \cite{Asboth:2005,Fernandez-Vidal:2010}). 

These features are consistent with the conjecture that selforganization is a second-order phase transition controlled by $\bar n$. This is also supported by the behaviour of the susceptibility, $\chi=\langle \Theta(t)^2\rangle-\langle |\Theta(t)|\rangle^2$, as a function of $\bar n$, which suggests a divergence at $\bar n_c$ for $N\to \infty$. We remark that the typical understanding of spatial domain formation at a second-order phase transition is here meaningless due to the non-additivity of the energy: mesoscopic Bragg gratings with $\Theta=\pm 1$ cannot stably coexist in space, since the resulting cavity field vanishes and with it the interatomic potential. 

\begin{widetext}
\begin{figure*}[hbt]
\begin{center}
\includegraphics[width=0.99\textwidth]{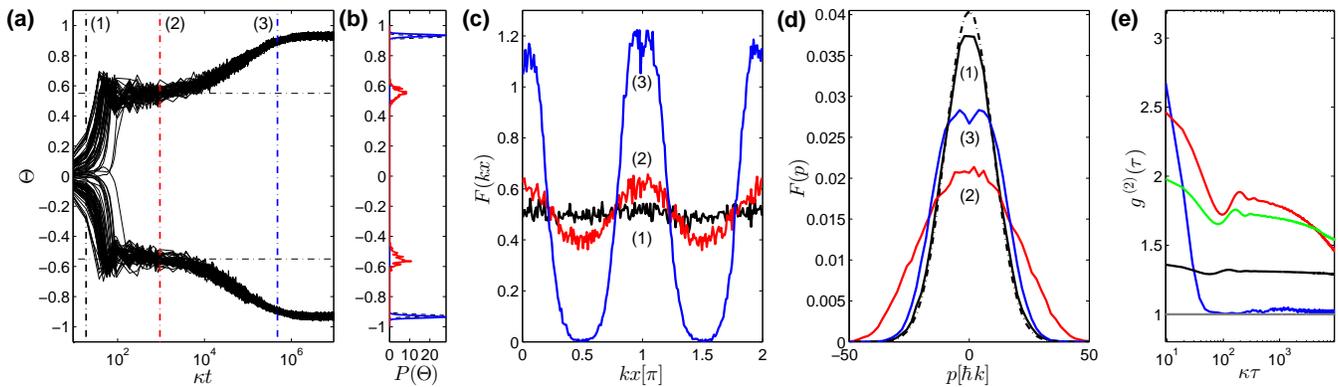}
\caption{\label{fig:3}  (color online) Dynamics of the order parameter above threshold: (a) $\Theta$ as a function of time (in units of $\kappa^{-1}$) for $N=200$ atoms and 500 trajectories at $\bar n=4\bar n_c$ and $\Delta_c=-\kappa$, for an initially spatially-uniform distribution at temperature $k_BT=\hbar \kappa/2$. $P(\Theta)$ at the transient and at the asymptotics is shown in panel (b). The position (c) and momentum (d) distributions are displayed at the times indicated by (1), (2), and (3) in panel (a). The dashed lines correspond to the initial distributions (which overlaps to (1) in (c)). (e) Intensity-intensity correlations of the light at the cavity output, $g^{(2)}(\tau)$ as a function of $\tau$ (in units of $\kappa^{-1}$) for $\bar n/\bar n_c=0.1,0.9,1,1.1,4$, (same color code as in Fig. \ref{fig:2}) evaluated after the system has reached the stationary state.}
\end{center}
\end{figure*}
\end{widetext}

We now turn to the dynamics leading to selforganization. We assume  that the initial distribution is spatially uniform, while the momentum distribution is a Maxwell-Boltzmann at width $\hbar\kappa/2$. For $\Delta_c=-\kappa$, at $\bar n \ll \bar{n}_c$ this distribution is stationary \cite{Schuetz:2013}. At t=0 the transverse field is quenched to a value corresponding to $\bar{n}$ above threshold. Figure \ref{fig:3} displays a sample of 500 trajectories of $\Theta(t)$  as a function of time when $\bar n = 4\bar n_c$ and $N=200$.  The trajectories are bunched and their behaviour can be ordered into three regimes, characterized by different time scales. First, a fast relaxation occurs over the time scale of dozens cavity lifetimes $\tau_c=1/\kappa$, in which the magnetization reaches an intermediate value of about 0.6 (Fig. \ref{fig:3}(b)), where it remains for a time scale exceeding $\tau_c$ by four orders of magnitude. During the relaxation the spatial density is almost uniform, therefore cross-correlations due to noise and mechanical forces are almost negligible. After this relaxation, part of the atoms form a Bragg grating (Fig. \ref{fig:3}(c)) while the momentum distribution is non-Gaussian (Fig. \ref{fig:3}(d)). We denote this regime by prethermalization. Then, the magnetization slowly grows  to the stationary value over time scales which are 6 orders of magnitude the cavity lifetime. Remarkably, for times of the order of $t\sim 10^5\tau_c$ the momentum distribution exhibits clear deviations from a Gaussian, and hence from a thermal state, even though the spatial distribution is very close to the asymptotic one. This behaviour can be understood considering that the diffusion is a function of the spatial distribution: As visible in the RHS of Eq. \eqref{FPE}, the strength of noise (and thus the relaxation rate) decreases the more the atoms are localized in the Bragg gratings, and thus at the nodes of the $\sin(kx)$ function. In the prethermalization time scale we verified that spatial diffusion follows a power law according to  $\langle x(t)^2\rangle\propto t^{2\alpha}$, where $\alpha$ is monotonously decreasing as $\bar n$ increases. In particular, it is superdiffusive ($\alpha>1/2$) below $\bar n_c$, while above $\bar n_c$ it becomes increasingly subdiffusive. In this latter case, in the long tails of relaxation it becomes normal again, $\alpha\to 1/2$. Figure \ref{fig:3}(e) displays $g^{(2)}(\tau)$ for different values of $\bar n$. Below threshold it rapidly decays from 3 to unity on a time scale of the order of cavity decay, at threshold its relaxation is orders of magnitude slower and exhibits damped oscillations, which can be associated with the density waves that become unstable and determine the Bragg grating (cif. Fig. \ref{fig:2}(d)). Well above threshold, instead, it remains locked to unity, corresponding to coherent light.  

The prethermalization behaviour, followed by the slow rate at which the steady state is approached, is typical above the selforganization threshold. We argue that it is a manifestation of the long-range correlations mediated by the cavity photons, and is analogous to observations made in studies of nonequilibrium stochastic long-range-interacting systems \cite{Bouchet:2012}. We further note that similar prethermalization features have been observed in quantum spin models with spatially-correlated noise \cite{Olmos:2012}. Differing from these latter models, here the stationary state exhibits long-range spatial correlations. On the other hand,  we do not find signatures of quasi-stationary states, whose relaxation times increase with $N^{\delta}$, with $\delta>1$ and whose existence is intrinsically related to the long-range nature of the interaction \cite{Campa:2009}. We believe this is due to the effect of the external environment, consistently with studies showing that its action can make these states dynamically unstable \cite{Gupta:2010,Chavanis:2011}.  

In this work we discarded the effect of spontaneous decay, assuming it is negligible as the laser field is far off resonance. Its role is expected to become more important as $\bar n$ is increased above threshold, and thus to enforce the dynamical instability of quasi-stationary states. Our model is also valid for any optically polarizable particles which can be confined within the resonator \cite{Vukics:2005}. It is also valid for $\bar n\gg\bar n_c$, when the atoms are tightly trapped in the potentials, as long as the effective trap frequency $\nu$ of the resulting lattice is smaller than the cavity linewidth \cite{Stenholm}. The description breaks down for $\nu\simeq \kappa$, when quantum mechanical coherence between the motional levels can be observed \cite{Wolke,Sandner}. 

In view of these results, one shall consider the selforganization transition observed in the ultracold regime by quenching the laser intensity \cite{Baumann:2010} in terms of an intrinsically out-of-equilibrium phenomenon. Indeed, our results predict that Hamiltonian solutions which possess the spatial modulation of the Bragg gratings will experience very small noise, even if they do not correspond with the stationary state. This raises the need to develop a kinetic theory for these systems as in Ref. \cite{Bouchet:2012}.  Preliminary studies in this direction have appeared in  \cite{Griessner:2010,Niedenzu:2012,DallaTorre:2013,Piazza:2014}. To conclude, our study shows that photonic systems offer a promising platform to study the statistical mechanics of long-range interacting systems, thus gaining insight into the dynamical properties of non-neutral plasmas and self-gravitating clusters \cite{Campa:2009}. 

The authors thank S. Fishman, H. Habibian, G. Manfredi, F. Piazza, H. Ritsch, L. Santen, W. Niedenzu, S. Ruffo, A. Vukics, and especially S. J\"ager for fruitful discussions. This work was partially supported by  the German Research Foundation (DFG). G.M. acknowledges the Ion Storage Group at NIST for hospitality, where part of this work was done.


\end{document}